\documentclass{iaus}
\usepackage{graphicx}
\usepackage{url}    % use the \url macro
\usepackage{natbib} % \citet \citep

\newcommand{\teff}{$T_{\rm{eff}}$}
\newcommand{\feh}{[Fe/H]}
\newcommand{\afe}{[$\alpha$/Fe]}

\title[~~Population models with $\alpha/Fe$] %% give here short title %%
{Spectral models of stellar populations resolved in chemical abundances}

\author[Philippe Prugniel \& Mina Koleva]   %% give here short author list %%
{Philippe Prugniel$^1$
 \and Mina Koleva$^{2,1}$}

\affiliation{$^1$Universit\'e de Lyon, Lyon I, CRAL-Observatoire de Lyon 
UMR5574, CNRS, France\\ email: {\tt philippe.prugniel@univ-lyon1.fr} \\[\affilskip]
$^2$Sterrenkundig Observatorium, Ghent University, Belgium}

\pubyear{2011} %% 
\volume{284}  %% insert here IAU Symposium No.
\pagerange{1--12}
\jname{The Spectral Energy Distribution of Galaxies}
\editors{R.J. Tuffs \&  C.C.Popescu, eds.}
\begin{document}

\maketitle

\begin{abstract}
We present model spectra of stellar populations with variable chemical 
composition.
We derived the [$\alpha$/Fe] abundance ratio of the stars of the 
most important libraries (ELODIE, CFLIB and MILES) using full spectrum
fitting and we generated PEGASE.HR models resolved in [$\alpha$/Fe].
We used a semi-empirical approach that combines the observed spectra
with synthetic stellar spectra.
We tested the models using them to derive  [$\alpha$/Fe] in 
galaxies and star clusters using full spectrum fitting.
The present models are available from
\url{http://ulyss.univ-lyon1.fr}


\end{abstract}

\firstsection % if your document starts with a section,
\section{Introduction}

The comparison between models and observations is the key to 
derive the metallicity, age and star formation history of galaxies
and star clusters.
The evolutionary synthesis of galaxies consists in choosing 
an initial mass function (IMF) and following the evolution of the
population over the time according to some theoretical evolutionary tracks.
This approach predicts the distribution
of the stars in the effective temperature (\teff) vs. surface gravity (log g)
plan at any epoch, and the final step is to draw spectra from a library
to assemble the integrated spectrum.

The modelling of stellar populations steadily progressed since
the first attempts in the 80s \citep[e.g.][]{bruzual1983}. 
Beside the metallicity, age and history, the detailed chemical
composition became an important parameter to measure.
This  allows one to trace the source of metal
enrichment, because of the different yields of type {\sc i}a 
and {\sc ii} supernovae. In turn, since both SN types correspond to different
time-scales, this informs about the duration of the
enrichment process. A population that produced its metal and
formed its stars before the  SN{\sc i}a onset will be deficient
in Fe (or as more commonly said, over-abundant in 
$\alpha$-elements, like O and Mg).
To vary the yield in the models, both the stellar library and the evolutionary tracks
have in principle to be changed.
However, \cite{dotter2007} have shown that the tracks essentially
depend on the total metallicity, rather than
of the actual mixture. Therefore, the scaled-solar tracks can reasonably be
used for any composition, provided they are associated with stellar spectra of the same 
total metallicity.

The models predict colours, used for SED fitting, spectra, and 
line-strength indices.
Indices are equivalent-widths defined for
a number of prominent spectral features and their study brings
an approximate handle on the physical parameters 
(age, metallicity and/or detailed composition).

Several authors modeled
indices with variable abundances of $\alpha$-elements relative to Fe
\citep{weiss1995, trager2000, thomas2003}.
The approach consisted in evaluating the response of individual indices
to changes of composition using theoretical spectra
\citep{tripicco1995}. Then, assuming that the abundance pattern in the library
is matching that of the solar neighborhood 
(justified by the fact that the observed libraries consist of nearby stars), 
the indices could be
estimated for any \teff{} log g, \feh{} and \afe.
In the visible region of the spectrum, the variation of \afe{}
results in a prominent change of the Mg$_b$ index near 5175~\AA, and therefore
a combination of this index with some Fe indices could be used to 
derive the \afe{} in galaxies and clusters.

Models of indices were developing faster and further than the spectral 
predictions, because of the lack of stellar libraries
calibrated in relative flux. 
However, the quality of the libraries also progressed.
Libraries of about 1000 stars covering a wide range of the
parameter space, with an extended coverage of the optical spectrum at mediun
spectral resolution and a fair flux calibration became available.
The three most important are ELODIE 
\citep[R=$\lambda/\Delta\lambda \approx$10000, $\lambda\lambda = 3900 - 6800$ \AA]{elodie31},
CFLIB \citep[R$\approx$4000, $\lambda\lambda = 3460 - 9400$ \AA]{valdes2004} 
and MILES \citep[R$\approx$2000, $\lambda\lambda = 3536 - 7410$ \AA]{sanchez-blazquez2006}.

In this paper, we present the first spectral models 
with variable \afe{}
based on these libraries and on precise measurements of \afe{}
in their stars.

\begin{figure}[b]
% \vspace*{-2.0 cm}
\begin{center}
 \includegraphics[width=0.95 \textwidth]{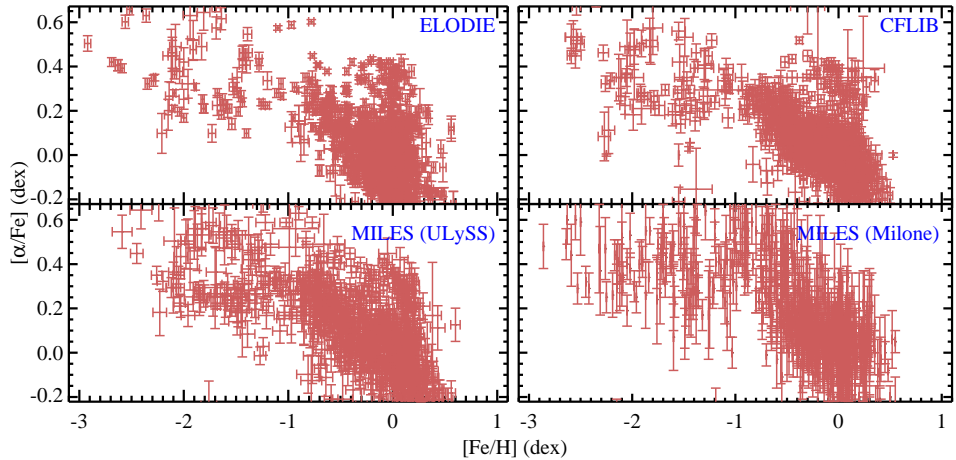} 
% \vspace*{-1.0 cm}
 \caption{Distribution of \afe{} in the three libraries, compared to the distribution 
in \cite{milone2011}.}
   \label{fig1}
\end{center}
\end{figure}

\section{Method}

Previous attempts to produce semi-empirical spectral models with variable \afe{}
were made in \citet{prugniel2007}, and \citet{walcher2009}.
The latter one uses population models based on synthetic spectra
to predict the differential change of each spectral element when
\afe{} is varied. This differential effect is then used to
`correct' empirical models to the desired $\alpha$-elements enrichment.
The former method determines the differential changes of the stellar spectra
and `correct' the library before computing models at different
\afe. Both approachs are equivalent, and are based on the assumption
that the libraries follow the abundance pattern of the solar 
neighborhood (i.e. they adopt a functional dependence of \afe{}
as a function of \feh). Models of Lick indices also make the same assumption.

Although the assumption of a relation between \afe{} and \feh{} is 
statistically verified, \afe{} present a significant dispersion
at a given \feh{} \citep[e.g.][]{milone2011}. 
When the library is interpolated to extract the
spectra entering in the synthesis, the stochastic effects can become 
significant. For example, for a local interpolation only a small number
of library stars are used near a given location in the \teff, log g and \feh{}
space, and their mean \afe{} may be biased. In order to avoid this effect, 
{\it the improvement introduced in the present work is to determine \afe{}
in the individual stars of the library.} 
Each stellar spectrum is then corrected to a set of given \afe{}
applying differential effects inferred from a theoretical library. 
An alternative may have been 
to measure \afe{} and directly interpolate the spectra in the four-dimensional
space. This would have avoided the dependency toward a theoretical library,
but required some extrapolations, because of the
non-uniform distribution of the stars in the 
\feh{} vs. \afe{} plan,, 

The approach is finally the following:
\begin{itemize}
\item[-] Measure \afe{} in the spectra of the libraries.
\item[-] Correct the library to some given values of \afe{} using
differential effects computed in a theoretical library.
\item[-] Build interpolators for these corrected libraries.
(these are functions returning a spectrum for given \teff, log g,
and \feh.
\item[-] Make model predictions at the chosen \afe{}.
\end{itemize}

\section{Determination of [$\alpha$/Fe] in the stars}

We used full-spectrum fitting with the ULySS package \citep{ulyss}
to measure \afe{}. Full-spectrum fitting was successful
to measure the other atmospheric parameters in the various libraries
\citep{wu2011,prugniel2011,koleva2011}, and the precision and
reliability was found consistent with detailed analysis using high-resolution
spectroscopy.

We measured \afe{} by comparing the
ELODIE spectra to the grid of \citet{coelho2005}. We minimized
the $\chi^2$ residuals to derive the four atmospheric parameters.
We measured \afe{} only for $3500 < T_{eff} < 7000$, because of
the limits of the grid of synthetic spectra. We assumed that the
remaining stars
follow the solar neighborhood abundance pattern.
Then, we used the ELODIE library as a reference to measure \afe{}
in the other libraries, using the same full-spectrum fitting method.

The measurements compare well with the most precise measurements from the
compilation of \citet{milone2011}. The distribution of \afe{} vs. \feh{} 
is presented in Fig.~\ref{fig1} for the three libraries and compared to
\citet{milone2011}.

\begin{figure}[b]
% \vspace*{-2.0 cm}
\begin{center}
 \includegraphics[width=0.97 \textwidth]{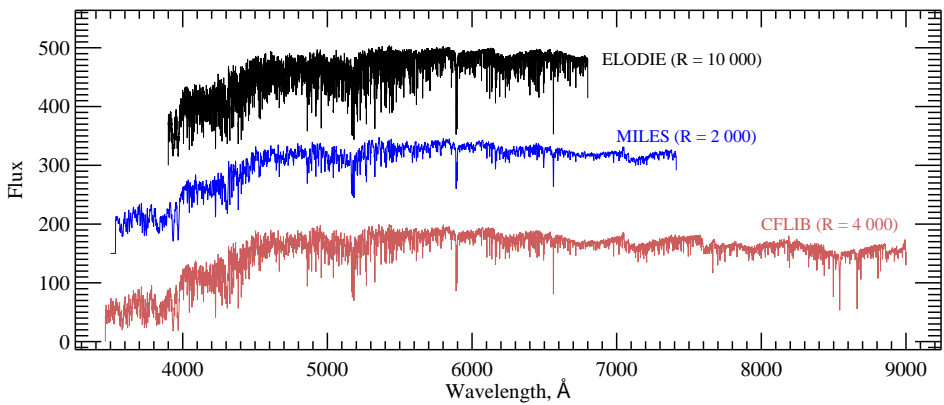} 
% \vspace*{-1.0 cm}
 \caption{Single stellar population models with PEGASE.HR and ELODIE, MILES and CFLIB. The age is eight Gyr and the metallicity is solar.}
   \label{fig2}
\end{center}
\end{figure}

\section{Population models resolved in [$\alpha$/Fe]}

We generated pairs of interpopators for \afe{} $\in\{0, 0.4\} $ dex,
and computed grids of single stellar 
populations (SSPs) models with PEGASE.HR \citep{leborgne2004}.
These pairs of grids were then grouped together to make four dimensional
arrays (wavelength, age, \feh{} and \afe) that can be used with ULySS
to determine the three SSP-equivalent parameters.
Figure~\ref{fig2} is a snapshot of models made with the three libraries.

We used these models to analyze Galactic globular clusters spectra as in \citet{koleva2008},
and we found results consistent with the determinations based on
spectroscopy of resolved stars. We also applied the method to study
spectra of galaxies from \citet{koleva2011b} and we found that the results
are consistent with those from Lick indices, but they are more precise.
The gain in precision reflects the better usage of the information. 
Mg$_b$ accounts for less than 20~\% of the
total opacity change in the optical range ($\lambda > 4500$~\AA) 
when \afe{} varies, and using the whole spectral range with full-spectrum 
fitting is a clear advantage.

\section{Prospective}
 
This semi-empirical method, used to build models of stellar populations
resolved in \afe, can straightforwardly be applied to resolve other
individual elements. The key point is to have reliable synthetic grids
of spectra to determine the differential effects.
The \cite{coelho2005} grid already
resolve Ca independently of the $\alpha$-elements, 
and other grids, varying other elements will be computed in the future.

The models presented here are available at 
\url{http://ulyss.uly-lyon1.fr}. They can be used with the ULySS
package to measure \afe{} in galaxies and clusters.

% If using bibtex is not allowed ... I can just take the .bst file
% ... for the moment assume that I can use 
\bibliographystyle{aa} % style aa.bst
\bibliography{prugniel}  % bibtex database

\end{document}